\documentclass[twocolumn,superscriptaddress,amsmath,amssymb,aps,prd,longbibliography]{revtex4-1}
\usepackage{dcolumn}
\usepackage{bm}
\usepackage{graphicx}
\usepackage{color}
\usepackage{hyperref}
\usepackage{amsmath}
\usepackage{multirow}
\usepackage{tabularx}
\usepackage{booktabs}
\usepackage{amsthm}

\newcommand{\be}{\begin{equation}}
\newcommand{\ee}{\end{equation}}
\newcommand{\ba}{\begin{eqnarray}}
\newcommand{\ea}{\end{eqnarray}}
\newcommand{\non}{\nonumber}
\newcommand{\n}[1]{\label{#1}}
\newcommand{\eq}[1]{(\ref{#1})}

\newcommand{\hhh}{\, ,\hspace{0.5cm}}
\newcommand{\BM}[1]{{\mbox{\boldmath $#1$}}}
\newcommand{\ind}[1]{\mbox{\tiny #1}}

\begin{document}

\title{Constraining mass of the graviton with GW170817}

\author{Andrey A. Shoom}
\email{andrey.shoom@aei.mpg.de}
\affiliation{Max Planck Institute for Gravitational Physics (Albert Einstein Institute), Leibniz Universität Hannover, Callinstr. 38, D-30167, Hannover, Germany}
\author{Sumit Kumar}
\email{sumit.kumar@aei.mpg.de}
\affiliation{Max Planck Institute for Gravitational Physics (Albert Einstein Institute), Leibniz Universität Hannover, Callinstr. 38, D-30167, Hannover, Germany}
\author{N. V. Krishnendu}
\email{krishnendu.naderi.varium@aei.mpg.de}
\affiliation{Max Planck Institute for Gravitational Physics (Albert Einstein Institute), Leibniz Universität Hannover, Callinstr. 38, D-30167, Hannover, Germany}


\begin{abstract}
We consider the massive graviton phenomenological model based on the graviton's dispersion terms included into phase of gravitational wave's waveform. Such model was already considered in many works but it was based on a single leading-order dispersion term only. Here we derive a relation between relativistic gravitons emission and absorption time intervals computed up to ${\cal O}(\gamma^{-6})$, where $\gamma$ is the Lorentz factor. Including the dispersion terms into the phase of gravitational wave's waveform results in two non-GR parameters of the $1^{\ind{-st}}$ and the $-2^{\ind{-nd}}$ post-Newtonian orders whose posteriors are used to put a constraint on the graviton's rest mass. We use the TaylorF2 waveform model to analyse the event GW170817 and report the following $95\%$-confidence upper bounds on the graviton's rest mass: $m^{\ind{Low Spin}}_{\ind{g}}\leq1.305\times10^{-54}$g and $m^{\ind{High Spin}}_{\ind{g}}\leq2.996\times10^{-54}$g for the high and low spin priors.
\end{abstract}

\maketitle

\section{Introduction}

The Einstein theory of gravity\textemdash general relativity (GR) is already more than hundred years old~\cite{Einstein:1916,Einstein:1918,EIH1938}. Since its first observational confirmation of the perihelion precession of Mercury's orbit and of the deflection of light by the Sun, it has been tested by many experiments and astronomical observations in the weak gravity scales \cite{Will:2014kxa}. However, the quest for going beyond the Einstein's GR never stopped. There are many proposed modifications of GR coming from assumptions of different origin, such as astronomical observations and theoretical models, e.g. modified Newtonian dynamics (MOND) \cite{Milgrom:1983ca,Milgrom:1983pn,Milgrom:1983zz} and bigravity models \cite{Hassan:2011vm,deRham:2010kj,Hassan:2011zd,Max:2017flc}. Here we shall pursue the idea of massive graviton.  

The idea of graviton\textemdash quanta of the gravitational field has a long story. The term {\em graviton} was coined by Blokhintsev and Gal'perin in 1934, in their paper on conservation of energy and neutrino hypothesis \cite{Gterm}. By 1949, successful quantisation of electromagnetic field by Feynman, Schwinger, and Tomonaga opened up a perspective to quantise the gravitational field. Yet the pursued ideas and attempts to successfully quantise the gravitational field have not brought us yet to a complete theory of quantum gravity \cite{smolin2008three,vidotto2015covariant}, and there are different points of view on the issue of gravitational field quantisation and different opinions were shared by Feynman \cite{Feynman} and Schwinger \cite{Schwinger}.  

In GR, gravitons are massless particles of spin 2 which propagate in vacuum with the speed of light. Massive gravity field theory was constructed by Ogievetsky and Polubarinov in \cite{OGIEVETSKY1965167} and much later was rediscovered by de Rham, Gabadadze, and Tolley \cite{deRham:2010kj} (see for comparison \cite{Mukohyama:2018bri}). Here we take phenomenological approach to massive graviton theory presented in \cite{Will:1997bb}. Namely, we consider massive graviton as a relativistic particle of mass $m_{\ind{g}}$ whose energy $E$ and 3-momentum $\BM{p}$ obey locally the relativistic energy-momentum relation $E^2=\BM{p}^2c^2+m_{\ind{g}}^2c^4$, where $c$ is the speed of light in vacuum. Using the expression for the graviton's 3-velocity $\BM{v}_{\ind{g}}=\BM{p}c^2/E$, the energy-momentum relation gives the dispersion relation
\be\n{intro1}
v_{\ind{g}}/c=\sqrt{1-m^2_{\ind{g}}c^4/E^2}\,,
\ee
which implies that more energetic gravitons move faster. Including this effect into phase of a gravitational wave (GW) from a compact binary coalescence should result in GW dispersion. The parameter which controls the dispersion effect is the graviton's mass. The leading-order dispersion term appears among the post-Newtonian (PN) expansion terms in GW waveform phase as a 1PN-order term where $m_{\ind{g}}$ is considered as an additional GW parameter \cite{Will:1997bb}. One can infer the {\em dynamic} upper bound value on the graviton's mass from its posterior distribution obtained from the parameter estimation (PE) runs. 

There is also {\em static} upper bound on $m_{\ind{g}}$ which is not related to propagation of gravitational interactions. Such a bound is based on Yukawa-type correction of the characteristic length scale $\sim m_{\ind{g}}^{-1}$ to the Newtonian gravitational potential, which is due to exchange of a massive mediator quanta\textemdash the massive graviton,
\be\n{intro2}
U(r)=-\frac{GM}{r}e^{-rm_{\ind{g}}c/h}\,.
\ee  
Here $h$ is the Planck constant. In this scenario, one searches for a rapid decay of a gravitational potential with a distance, which indicates the Yukawa exponential cutoff. The first estimate of the graviton mass based on analysis of bound clusters of galaxies was reported in \cite{Goldhaber:1974wg}. Taking 580kpc as the maximum separation of galaxies in clusters for the distance over which gravity decreases by the factor of $e^{-1}$ the rest mass of graviton was found to be less than $2\times10^{-62}$g. However, the galaxies cluster bound is a crude estimate and it may also be explained by the presence of dark matter. Therefore, this estimate may not be reliable. Better estimates of the graviton's Compton wavelength $\lambda_{\ind{g}}=h/m_{\ind{g}}c$ were based on solar system data analysis under Yukawa gravitational potential assumption and verification of Kepler's third law for the inner planets \cite{Talmadge:1988qz,Will:1997bb}. The reported value for gravitational coupling $\geq 10^{15}$m gives the upper bound on the graviton's mass $m_{\ind{g}}\leq 2.2\times 10^{-54}$g. 

The first dynamic bound of $m_{\ind{g}}<7.6\times10^{-20}$eV/c$^{2}=1.36\times 10^{-52}$g with 90\% confidence was found from analysis of the observed orbital decay of the binary pulsars PSR B1913+16 and PSR B1534+12 \cite{Finn:2001qi}. GW150914 data was used to put the dynamic upper bound $m_{\ind{g}}\leq 2.14\times 10^{-55}$g with 90\% confidence \cite{TheLIGOScientific:2016src}. There is also another phenomenological model introduced in \cite{PhysRevD.85.024041}. This model is based on a Lorenz-violating dispersion relation 
\be\n{intro3}
E^2=p^2c^2+m^2_{\ind{g}}c^4+Ap^{\alpha}c^{\alpha}\,,
\ee
where $A$ and $\alpha$ are the Lorentz-violating parameters. The first application of this model (though in the trivial case of $A=\alpha=0$) to LIGO data of GW event GW170104 was reported in \cite{PhysRevLett.118.221101,PhysRevLett.121.129901}, where the combined lower bound was found to be $\lambda_{\ind{g}}>1.6\times10^{16}$m, which corresponds to the graviton's mass bound $m_{\ind{g}}\leq1.37\times10^{-55}$g. The same choice of the parameters for GW170817 event gave the value $m_{\ind{g}}\leq1.70\times10^{-54}$g \cite{Abbott:2018lct}. For a set of different values of $\alpha$ and $A$ parameters the combined from ten GW events the bound $m_{\ind{g}}\leq 4.7\times10^{-23}$eV/$c^2=8.4\times10^{-56}$g was reported in \cite{PhysRevD.100.104036}. 

In this paper we restrict our attention to the binary neutron star merger GW170817 event 
\cite{TheLIGOScientific:2017qsa}. For this event, the contribution of the merger to signal-to-noise-ratio is not significant, whence the inspiral regime and standard PN methods dominate \cite{Blanchet:1995ez,Blanchet:2001ax,Blanchet:2004ek,Blanchet:2008je,Blanchet:2013haa}. Moreover, the source proximity allows us to neglect the cosmological redshift, which simplifies our model. We consider the standard (not Lorentz-violating) massive graviton dispersion relation \eq{intro1}. It could be that gravitons exist only as massless particles, quanta of the gravitational field, or it could also be that gravitons do not exists in nature, i.e. there are no mediators of gravitational interaction, or more strongly, gravity cannot be quantised. In that case ``massive gravitons" would imply a modification of general relativity from which follows a nontrivial dispersion relation of (classical) GWs in the linear (weak gravity) regime. In such a scenario ``graviton's mass" might be considered as a GW dispersion parameter. For instance, modification of GW dynamics, such as GW birefringence, was considered and tested on GW events in \cite{Wang:2021gqm}. The goal of our work is to use the current GWs observational data and put constraints on graviton's mass (or the GW dispersion parameter). 

Our paper is organised as follows. In Sec.~\ref{sec:kinematics} we present the massive graviton kinematics and derive a relation between emission and absorption time intervals computed from the first two terms in the inverse Lorentz factor expansion of the massive graviton dispersion relation. In Sec.~\ref{sec:massive_gravitation_phase} using the TaylorF2 waveform model we apply the time intervals relation to compute phase of GW waveform. This results in two non-GR parameters of 1PN and -2PN orders. Section~\ref{sec:massive_gravitation_results} contains results and their analysis for GW170817 event. In conclusion we summarise our results and discuss future prospectives of related work. 

\section{Kinematics of the massive graviton}\label{sec:kinematics}

\begin{figure}[htb]
\begin{center}
\includegraphics[width=5cm]{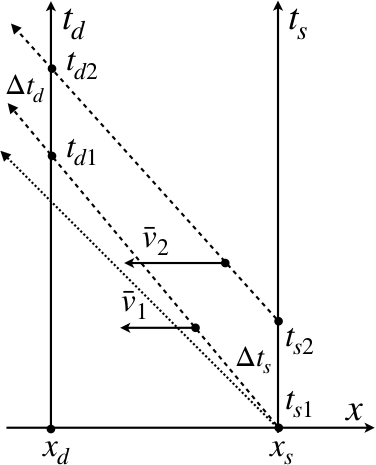}
\caption{Propagation of massive gravitons of different energies from their source $(t_{s}, x_{s})$ to the detector $(t_{d}, x_{d})$ in Minkowski space-time. The source and the detector are mutually at rest and their clocks measuring $t_{s}$ and $t_{d}$ are synchronised. The dotted line represents world line of a photon emitted at $(t_{s1}, x_{s})$ toward the detector. The dashed lines represent world lines of massive gravitons.}\label{fig1}
\end{center}
\end{figure}

As it follows from the dispersion relation \eq{intro1}, more energetic gravitons move faster. We are interested in finding a relation between time intervals of emission and detection of such gravitons. Consider the following situation: source of massive gravitons and their detector are mutually at rest and located in Minkowski space-time. Let a graviton of energy $E_{1}$ is emitted at time $t_{s1}$ measured by the source clock. The graviton moves toward the detector with 3-velocity $\BM{v}_{1}$ and arrives at time $t_{d1}$ measured by the detector clock, which is synchronised with the source clock. Let another graviton of energy $E_{2}$ is emitted at time $t_{s2}$. It moves toward the detector with 3-velocity $\BM{v}_{2}$ and arrives at time $t_{d2}$. Space-time diagram illustrating this process is shown in Fig.~\ref{fig1}.

\noindent 
Then, from the kinematic relation
\be\n{2.1}
D=x_{s}-x_{d}=v_{1}(t_{d1}-t_{s1})=v_{2}(t_{d2}-t_{s2})\,,
\ee
where $D$ is the proper distance between the source and the detector, we derive the following relation between the time intervals $\Delta t_{s}=t_{s2}-t_{s1}$ and $\Delta t_{d}=t_{d2}-t_{d1}$:
\be\n{2.2}
\Delta t_{d}=\Delta t_{s}-D(v_{1}^{-1}-v_{2}^{-1})\,.
\ee
Using the dispersion relation \eq{intro1} and assuming that the gravitons are highly relativistic, that is, $\gamma = E/m_{\ind{g}}\gg 1$, we derive
\be\n{2.3}
\left.v_{g}^{-1}\right|_{\gamma\ll1}=1+\frac{m_{\ind{g}}^2}{2E^2}+\frac{3m_{\ind{g}}^4}{8E^4}+{\cal O}(\gamma^{-6})\,.
\ee
Then, using the graviton wave-particle duality relation $E=hf$, where $h=6.62607015\times10^{-34}$J$\cdot$s and $f$ is the graviton's frequency, we derive
\be\n{2.4}
\Delta t_d\approx\Delta t_s-\left[\frac{Dm_{\ind{g}}^2}{2h^2}(f_{1s}^{-2}-f_{2s}^{-2})+\frac{3Dm_{\ind{g}}^4}{8h^4}(f_{1s}^{-4}-f_{2s}^{-4})\right]\,.
\ee
Note that this relation has to be modified if the source and the detector are in a relative motion or located sufficiently far, so that one has to take into account cosmological redshift. The measured luminosity distance for the event GW170817 is about 40Mpc, and the corresponding cosmological redshift is $z\approx0.008$ \cite{TheLIGOScientific:2017qsa}. Thus, we can neglect the cosmological redshift and use the expression above \footnote{We derived the general expression taking into account the cosmological redshift factor and found that for $z=0.008$ the maximal relative error in the final expression for the graviton's mass is about 0.0006.}.

\section{Phase of GW in the massive graviton scenario}
\label{sec:massive_gravitation_phase}

In this section, we consider GW from an inspiralling binary system. We describe dynamics of the binary in the binary frame with the origin at its original centre of mass location and the binary evolution time $t_b$ \footnote{In addition to energy and angular momentum, GWs carry linear momentum too. Therefore, an inspiralling binary system acquires recoil velocity and as a result, its centre of mass moves away from its original position (for a recent analysis of the gravitational recoil see e.g. \cite{PhysRevLett.124.101104})}. Here and in what follows, the subscript $b$ stands for the binary frame. Assuming that the binary orbit evolves adiabatically, i.e. its orbital angular velocity change over the orbital period $T_b$ is very small, the binary energy rate of change is also small. Then, the following dynamic equations hold approximately:
\ba
&&\frac{d\phi(t_b)}{dt_b}=\frac{v^3}{M_b}\,,\n{3.1a}\\
&&\frac{dv}{dt_b}=-\frac{{\cal F}(v)}{M_b{\cal E}'(v)}\,.\n{3.1b}
\ea
Here $\phi(t_b)$ is the binary phase (azimuthal angle of the binary reduced mass), $M_b$ is the total mass of the binary,
\be\n{3.2}
v=(\pi M_b f_b)^{1/3}\ll 1\,,
\ee
is the typical speed of the binary circular motion, where $f_b=2/T_b$ is the radiated gravitational wave frequency, ${\cal F}(v)$ is the GW power output (GW luminosity), ${\cal E}(v)$ is the dimensionless total mechanical energy of the binary, and ${\cal E}'=d{\cal E}/dv$. The first equation \eq{3.1a} is the Kepler's third law and the second one \eq{3.1b} is the energy balance equation. 

We shall need to know the binary evolution time $t_b$ as a function of $f_b$. This can be derived by using \eq{3.2} and integrating \eq{3.1b},
\be\n{3.3}
t_b(f_b)=t_c-\frac{\pi M_b^2}{3}\int^{f_c}_{f_b}\frac{{\cal E}'(v)}{{\cal F}(v)}\frac{df_b}{v^2}\,,
\ee
where $f_c$ is the coalescence frequency corresponding to coalescence time $t_c$ measured in the binary frame. To compute this integral we need to know the functions ${\cal E}(v)$ and ${\cal F}(v)$. They are given in terms of power series in $v$ up to 3.5PN order ($\sim(v/c)^7$) \cite{Blanchet:1995ez,Faye:2012we,Buonanno:2009zt,Bohe:2015ana,Bohe:2013cla}. Then one can compute their ratio and expand it in powers of $v$ up to the 3.5PN order. 

Gravitational wave radiated by the binary motion is observed at the detector frame. We consider GW waveform model of the following general form, without specifying its polarisation modes:
\be\n{3.4}
h(t_d)=A(t_d)\cos\varphi(t_d)\,.
\ee
Here $t_d$ is time measured in the detector frame, $A(t_d)$ is the GW amplitude and $\varphi(t_d)$ is its phase. Here and in what follows, the subscript $d$ stands for the detector frame. The Fourier transform of $h(t_d)$ is defined as
\be\n{3.5}
\tilde{h}(\tilde{f})\equiv\int^{+\infty}_{-\infty}h(t_d)e^{i2\pi\tilde{f}t_d}dt_d\,.
\ee
For the given problem it is rather impossible to compute the Fourier transform analytically and we shall use the stationary phase approximation based on the condition $\tilde{f}\gg1$ (see, e.g. \cite{Bender-Orszag} and also \cite{Cutler:1994ys}). Integration of \eq{3.5} by parts and requirement that the derived integral is negligible as compared to the boundary term impose the following conditions on its integrant:
\be\n{3.6}
\frac{d\ln A(t_d)}{dt_d}\ll\frac{d\varphi(t_d)}{dt_d}\hhh \frac{d^2\varphi(t_d)}{dt_d^2}\ll\left(\frac{d\varphi(t_d)}{dt_d}\right)^2\,.
\ee
These conditions are fulfilled for astrophysical binary systems during their inspiral and up to their coalescence (see, e.g. \cite{Cutler:1994ys}). Then, the method of stationary phase yields the following leading asymptotic behaviour of the Fourier transform:
\be\n{3.7}
\tilde{h}(f_d)\approx\frac{1}{2}A(f_d)\left(\frac{dt_d}{df_d}\right)^{1/2}e^{i(\Psi(f_d)-\pi/4)} \,.
\ee
Here $t'_d$ is a solution to the the stationary point equation
\be\n{3.7a}
\frac{d\varphi(t'_d)}{dt_d}=2\pi f_d(t'_{d})=2\pi\tilde{f}\,,
\ee
where $df_d(t'_d)/dt_d\ne0$ and
\be\n{3.8} 
\Psi(f_d)=2\pi\int^{f_c}_{f_d}[t_c-t_d(f_d)]df_d+2\pi t_cf_d-\varphi_c\,.
\ee 

To evaluate the integral we need to know the function $t_d(f_d)$. The expressions \eq{3.2} and \eq{3.3} define the function $t_b(f_b)$. Frequencies $f_b$ and $f_d$ are related through the redshift factor, which we neglect here, because as it was said in the previous section, for the event GW170817 it equals to $1.008\approx1.0$. Thus, we consider 
\be\n{3.9}
f_b\approx f_d\equiv f\,.
\ee
Time intervals in the binary and the detector frame are related via \eq{2.4}. Defining 
\ba\n{3.10}
\Delta t_d=t_c-t_d(f_d)&\hhh&\Delta t_s=t_c-t_b(f_b)\,,\non\\
f_{1s}=f&\hhh&f_{2s}=f_c\,,
\ea
we derive
\ba\n{3.11}
t_c-t_d(f_d)&\approx&t_c-t_b(f)\\
&-&\frac{Dm_{\ind{g}}^2}{2h^2}(f^{-2}-f_{c}^{-2})-\frac{3Dm_{\ind{g}}^4}{8h^4}(f^{-4}-f_{c}^{-4})\,.\non
\ea
Integrating the first term in \eq{3.11} gives TaylorF2 approximation of the GW phase (see, e.g. \cite{Buonanno:2009zt}), 
\be\n{3.12}
\Psi^{\ind{GR}}(f)=\sum^{N}_{n=0}\left(\frac{f}{f_{\ind{ref}}}\right)^{(n-5)/3}\left[\Psi_{n}+\Psi^{\text{ln}}_{n}\ln\left(\frac{f}{f_{\ind{ref}}}\right)\right]\,,
\ee
where we use the reference frequency $f_{\ind{ref}}=1/\pi M$, where $M=(z+1)M_b\approx M_b$. In the series the only non-zero logarithmic terms are $\Psi^{\ln}_{5}$ and $\Psi^{\ln}_{6}$. The expansion of  order $N$ corresponds to $(N/2)$PN order. In this expansion the 2.5PN term becomes indistinguishable from the binary coalescence phase term $(-\varphi_c-\pi/4)$ and the 4PN term is indistinguishable from the binary coalescence time term $2\pi f\,t_c$. For GW170817 the PN terms $\Psi_{n}$ are functions of the binary physical parameters: masses of the stars, their orbit-aligned dimensionless spin components, and their tidal deformability parameters.

Integration of the second and third terms in \eq{3.11} gives us non-GR terms due to the massive graviton,
\be\n{3.15}
\delta\Psi(f)=\delta\Psi_{2}\left(\frac{f}{f_{\ind{ref}}}\right)^{-1}+\delta\Psi_{-4}\left(\frac{f}{f_{\ind{ref}}}\right)^{-3}\,,
\ee
where
\be\n{3.16}
\delta\Psi_{2}=-\frac{\pi^2MDm_{\ind{g}}^2}{h^2}\hhh \delta\Psi_{-4}=-\frac{\pi^4M^3Dm_{\ind{g}}^4}{4h^4}
\ee
are terms of the 1PN and $-2$PN order. Thus, the phase \eq{3.8} now reads
\be\n{3.17}
\Psi(f)=\Psi^{\ind{GR}}(f)+\delta\Psi(f)\,.
\ee

Note that the gravitational wave amplitude ${\cal A}(f)$ can also be expanded as a post-Newtonian series, but here we will keep only terms at Newtonian order and hence ${\cal A}(f) \sim f^{-7/6}$. 

Finally, we discuss PN order $(N/2)$ in the phase expansion \eq{3.12} required for consistent measurement of the graviton's mass. As we already mentioned in the introduction, massive graviton model implies a Yukawa-type potential \eq{intro1}. This potential can be derived from the propagator of a massive meson (in our case it is the massive graviton), which mediates interaction between two fermions (in our case between massive objects). The gravitational force corresponding to the potential \eq{intro2} is 
\be\n{3.18}
\BM{F}=-\nabla U(r)=-\frac{M\hat{\BM{r}}}{r^2}\left(1+\frac{r}{\lambda_{\ind{g}}}\right)e^{-r/\lambda_{\ind{g}}}\,,
\ee
where $\lambda_{\ind{g}}=h/m_{\ind{g}}c$ is Compton wavelength of the massive graviton and $\hat{\BM{r}}$ is a unit radial vector. It is typical for a compact binary motion that the characteristic range of gravitational interaction is much less than the estimated value of $\lambda_{\ind{g}}\sim10^{15}$m \cite{Talmadge:1988qz,Will:1997bb}. Thus we can expand \eq{3.18} in powers of $r/\lambda_{\ind{g}}$,
\be\n{3.19}
\left.\BM{F}\right|_{r\ll\lambda_{\ind{g}}}\approx-\frac{M\hat{\BM{r}}}{r^2}\left(1-\frac{r^2}{2\lambda_{\ind{g}}^2}\right)\,.
\ee
If we now assume that $r^2/\lambda_{\ind{g}}^2\ll v^N$ holds for $N$ in \eq{3.12} then we can neglect the massive graviton effects on the binary dynamics. As it follows from the Kepler's law \eq{3.1a}, for a nearly circular binary orbit $r\approx M/v^2$, while for a neutron star binary $r<6M$, thus $v<1/\sqrt{6}$. Then we should have $M\ll 6^{-N/4-1}\times10^{15}$m. This inequality clearly holds for the binary mass $M\sim M_{\odot}\approx10^3$m and $N=7$. Therefore, one can safely neglect the massive graviton's contribution to binary dynamics for 3.5PN order. 

\section{Results and analysis}
\label{sec:massive_gravitation_results}

\begin{figure}[htb]
\begin{center}
\includegraphics[width=7cm]{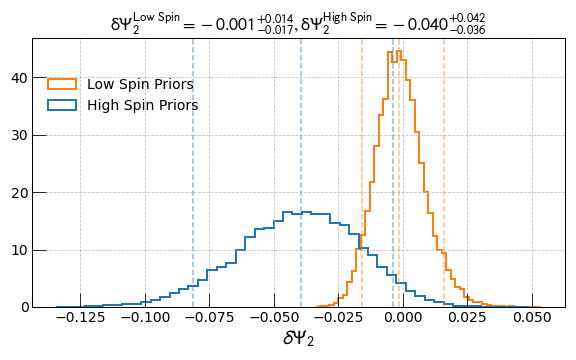}
\caption{Marginalized 1D posterior distribution for non-GR parameter $\delta\Psi_{2}$ for low and high spin priors of the binary components from the parameter estimation run where only the leading order non-GR parameter was considered.}\label{fig2}
\end{center}
\end{figure} 
\begin{figure*}[htb]
\begin{center}
\includegraphics[width=10cm]{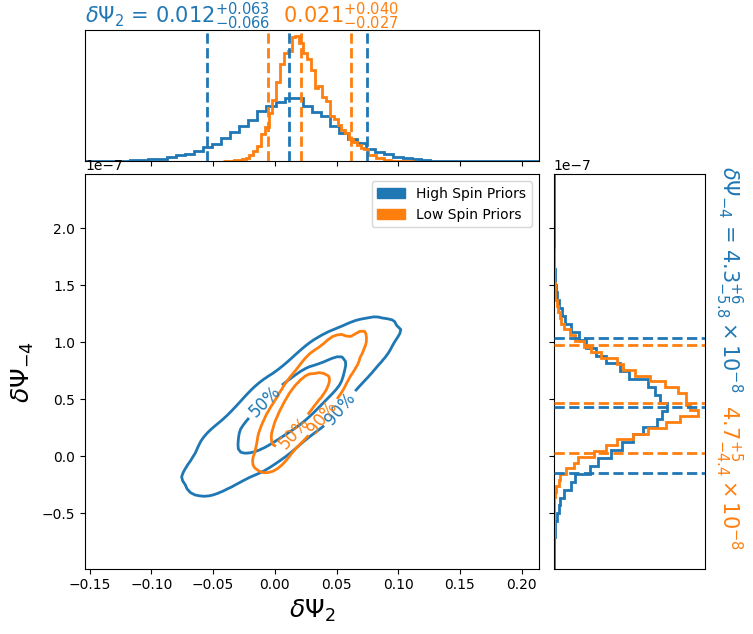}
\caption{Marginalized 2D posterior distribution of the non-GR parameters $\delta\Psi_{2}$ and $\delta\Psi_{-4}$ for low and high spin priors of the binary components.}\label{fig3}
\end{center}
\end{figure*} 
\begin{figure}[htb]
\begin{center}
\includegraphics[width=8cm]{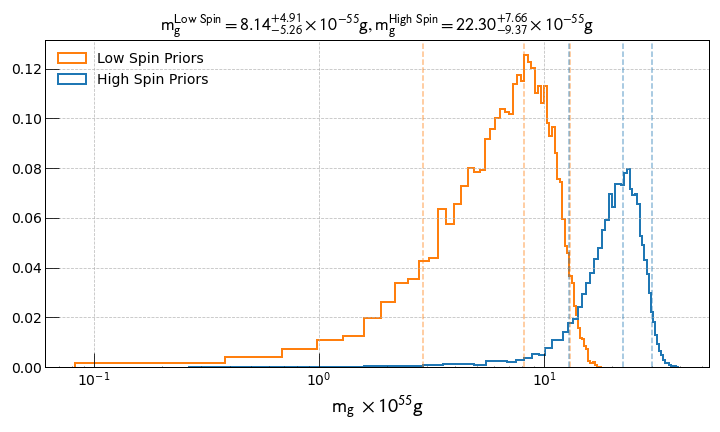}
\caption{Proper mass of the massive graviton posterior from the parameter estimation run where we consider one non-GR parameter $\delta\Psi_{2}$ alone for low and high spin priors of the binary components. The distribution is constructed from the posterior of $\delta\Psi_{2}$ with the prior range $[-5, 0]$, which in accordance with \eq{3.16}.}\label{fig4}
\end{center}
\end{figure} 
\begin{figure*}[htb]
\begin{center}
\ba
&&\hspace{0cm}\includegraphics[width=8cm]{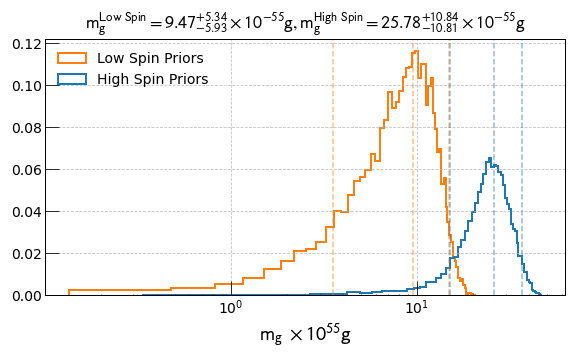}
\hspace{0.5cm}\includegraphics[width=8cm]{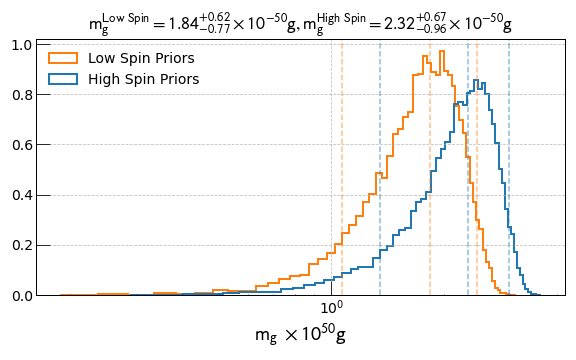}\non\\
&&\hspace{4.0cm}({\bf a})\hspace{8.0cm}({\bf b})\non
\ea
\caption{Proper mass of the massive graviton posterior from parameter estimation run where we vary both the non-GR parameters $\delta\Psi_{2}$ and $\delta\Psi_{-4}$ for low and high spin priors of the binary components. Plot {\bf (a)} illustrates the distribution constructed from the marginalized 1D posterior of $\delta\Psi_{2}$ and plot {\bf (b)} illustrates the distribution constructed from the posterior of $\delta\Psi_{-4}$. For these PE runs, we use the uniform prior for non-GR parameters and restrict them to the negative values: $\delta\Psi_{2} \in [-5,0], \delta\Psi_{-4} \in [-10^{-5},0]$ in accordance with \eq{3.16}.}\label{fig5}
\end{center}
\end{figure*}

According to the Bayes's theorem, the posterior distribution $p(\theta|D,I)$ for the parameters $\theta$ of a model, in the light of observed data $D$, and prior knowledge $I$ is given by,
\begin{equation}
p(\theta|D,I) = \frac{\mathcal{L}(D|\theta,I)\pi(\theta|I)}{p(D|I)},
\end{equation}
\noindent
where $\mathcal{L}(D|\theta,I)$ is the likelihood function, $\pi(\theta|I)$ is the prior probability function on the parameters $\theta$, and $p(D|I)$ is known as Bayesian evidence or marginalized likelihood. We can ignore the term in denominator if we are interested only in the posterior probability distribution for the given model, as this term contributes to the normalization factor. Assuming detector noise to be Gaussian and stationary around the event, we make use of standard likelihood function for the residual $\tilde{s}(\tilde{f})-\tilde{h}(\tilde{f})$,  where $\tilde{s}(\tilde{f})$ is Fourier transform of the GW strain $s(t)$, and $\tilde{h}(\tilde{f})$ is the Fourier transform of the waveform template $h(t)$. To obtain the posteriors of our model parameters, we make use of the publicly available PyCBC inference package \cite{Biwer:2018osg}. We use TaylorF2 waveform model \cite{Blanchet:1995ez, Faye:2012we} to generate the waveforms, implemented in \textsc{LALSuite} \cite{lalsuite}. We make use of heterodyne likelihood model described in ~\cite{Cornish:2010kf,Finstad:2020sok,Zackay:2018qdy}. We use the flat priors on source frame masses, comoving volume, trigger time $t_c$, tidal deformability parameters $\Lambda_1,\Lambda_2 \in [0,5000]$, and testing GR parameters $\delta\Psi_2 \in [-5,5]$ and $\delta\Psi_{-4} \in [-10^{-5},10^{-5}]$.
We use isotropic priors on polarization, right ascension, declination, inclination angle, and spins. We use aligned spin model with priors projected along the spinning axes from the isotropic spin priors. Furthermore, we use two prior ranges for spins: i) low spin priors where we restrict the magnitude of the isotropic spins $|s_{1a,2a}| \in[0,0.05]$, and ii) high spin priors where we allow the magnitude of the spins to vary in wider range $|s_{1a,2a}| \in[0,0.5]$. For sampling, we use the publicly available implementation of nested sampling sampler \textsc{dynesty} \citep{speagle:2019}.

We modify the GR waveform for the GW170817 event by adding the first non-GR parameter $\delta\Psi_{2}$ and then both the non-GR parameters $\delta\Psi_{2}$ and $\delta\Psi_{-4}$. 

\begin{table*}[htb]
\caption{Constraints on the mass of the graviton from various parameter estimation runs. Here we report the $95\%$ upper bound on the $m_{\ind{g}}$ distribution constructed from 1D marginalized posteriors for non-GR parameters as shown in figures \ref{fig4} and \ref{fig5}.\newline}
\label{tab:constraints_mg}
\begin{tabular}{|c|c|cc|}
\hline
\multirow{2}{*}{Varying non-GR Parameter(s)} & \multirow{2}{*}{\begin{tabular}[c]{@{}c@{}}Parameter Used to \\ Constrain $m_{\ind{g}}$ \end{tabular}} & \multicolumn{2}{c|}{\begin{tabular}[c]{@{}c@{}}Upper Bounds on $m_{\ind{g}}$\\ (in g)\end{tabular}} \\ \cline{3-4} 
                                          &                                                                                                     & \multicolumn{1}{c|}{Low Spin Priors}                       & High Spin Priors                     \\ \hline
$\delta\Psi_2$                            & $\delta\Psi_2$                                                                                      & \multicolumn{1}{c|}{$1.305 \times 10^{-54}$}               & $2.996 \times 10^{-54}$              \\ \hline
$\delta\Psi_2, \delta\Psi_{-4}$           & $\delta\Psi_2$                                                                                      & \multicolumn{1}{c|}{$1.481 \times 10^{-54}$}               & $3.661 \times 10^{-54}$              \\ \hline
$\delta\Psi_2, \delta\Psi_{-4}$           & $\delta\Psi_{-4}$                                                                                   & \multicolumn{1}{c|}{$2.45 \times 10^{-50}$}                & $2.98 \times 10^{-50}$               \\ \hline
\end{tabular}
\end{table*}

We perform various combination of parameter estimation runs: GR run, non-GR run with leading order non-GR term $\delta\Psi_{2}$, and non-GR run varying both the parameters $\delta\Psi_{2}$ and $\delta\Psi_{-4}$. We perform all these runs for low spins and high spins priors. In Fig.~\ref{fig2}, we show the marginalized one-dimensional posterior distribution for $\delta\Psi_{2}$ parameter from the runs where only leading order non-GR term was taken along with GR parameters. Fig.~\ref{fig3} shows the 2D marginalized posterior distribution from the runs where both non-GR parameters are varied along with GR parameters.

Using results of our runs we construct the covariance matrix of the non-GR parameters. For low spin priors it is
We perform various combination of parameter estimation runs: GR run, non-GR run with leading order non-GR term $\delta\Psi_{2}$, and non-GR run varying both the parameters $\delta\Psi_{2}$ and $\delta\Psi_{-4}$. We perform all these runs for low spins and high spins priors. In Fig.~\ref{fig2}, we show the marginalized one-dimensional posterior distribution for $\delta\Psi_{2}$ parameter from the runs where only leading order non-GR term was taken along with GR parameters. Fig.~\ref{fig3} shows the 2D marginalized posterior distribution from the runs where both non-GR parameters are varied along with GR parameters. 

Using the expression \eq{3.16} we can estimate the graviton's proper mass as follows:
\be\n{4.4a}
m^{\text{1PN}}_{\ind{g}}\approx\frac{h}{\pi\,c}\left(\frac{-\delta\Psi_{2}}{MD_{L}}\right)^{1/2}
\ee
from the 1PN order and 
\be\n{4.4b}
m^{\text{-2PN}}_{\ind{g}}\approx\frac{h}{\pi\,c}\left(\frac{-4\,\delta\Psi_{-4}}{M^3D_{L}}\right)^{1/4}
\ee
from the $-2$PN order. Here we take $D=D_L$\textemdash the luminosity distance measured in Mpc, where $1\mbox{pc}=3.0857\times10^{16}$m, the speed of light $c=299792458$m/s, and we measure the total mass $M$ in the units of geometrized solar mass $M_{\odot}=1.48\times10^3$m. 

Using these expressions we can derive posteriors of $m_{\ind{g}}$ from the posteriors of the non-GR parameters. It can be seen from the equation \eq{3.16} that only negative values of non-GR parameters are valid for the expression of the graviton mass, therefore, to put constraints on $m_{\ind{g}}$, we perform another set of PE runs where we restrict the prior range to negative values i.e. $\delta\Psi_2 \in [-5,0], \delta\Psi_{-4} \in [-10^{-5},0]$.

The Fig.~\ref{fig4} plot shows posterior distribution of $m_{\ind{g}}$ from $\delta\Psi_{2}$ alone for low and high spin priors of the binary components. The Fig.~\ref{fig5} plots show posterior distribution of $m_{\ind{g}}$ from $\delta\Psi_{2}$ and $\delta\Psi_{-4}$ for low and high spin priors of the binary components. The $95\%$ upper bounds on the graviton's mass are summarised in Table~\ref{tab:constraints_mg}. Combining these results we conclude that the upper bound on the graviton mass is
\ba
m^{\ind{Low Spin}}_{\ind{g}}&\leq&1.305\times10^{-54}\mbox{g}\,,\n{4.8a}\\
m^{\ind{High Spin}}_{\ind{g}}&\leq&2.996\times10^{-54}\mbox{g}\,.\n{4.8b} 
\ea
To finalise our analysis we would like to mention first that including the second non-GR parameter $\delta\Psi_{-4}$ does change the posterior of the first (the dominant) one $\delta\Psi_{2}$ due to their mutual correlation. This can be seen directly from the figures \ref{fig2} and \ref{fig3}. Second, the graviton mass upper bound derived from $\delta\Psi_{-4}$ posterior is of four orders of magnitude larger, as compared to the upper bound derived from $\delta\Psi_{2}$. This is likely due to relatively weak contribution of the non-GR term $\delta\Psi_{-4}$ to the waveform phase. Such a weak contribution is also largely contaminated with noise and, as a result, has less constraining power as compared to the first non-GR term.   

\section{Conclusion}
\label{sec:massive_gravitation_con}

The upper bound estimates on the graviton mass found here [\eq{4.8a} and \eq{4.8b}] are of the same order of magnitude as the previously reported value $m_{\ind{g}}\leq1.70\times10^{-54}$g \cite{Abbott:2018lct}. The main goal of this work was to include the subleading-order term in the dispersion relation expansion of massive graviton into GW's waveform phase. This term alone gives a few orders of magnitude ($m_{\ind{g}}\sim 10^{-50}$g) larger upper bound estimate than the leading one. This is expected due to relatively weak contribution of the non-GR term $\delta\Psi_{-4}$ to the waveform phase. Yet, the inclusion of this term does affect the leading non-GR term posterior due to their mutual correlation. Moreover, one is motivated to explore the phenomenological model of the massive graviton in detail, by considering the sub-leading dispersion terms as well. This is the approach we developed and explored in this work. We expect that with the advert of more sensitive future techniques more accurate estimates on the subleading non-GR term(s) can be derived. The formalism presented here can naturally be extended by taking into account background space-time curvature, that allows to include cosmological models and test the massive graviton model on other GW events from remote sources which we plan to consider in our future research. Finally, we would like to mention that the posteriors of the non-GR parameters confirm validity of GR, yet one may naturally expect its extension with new gravity models which encompass GR at certain gravity scale. 

\begin{acknowledgments}

The computation of the work was run on the ATLAS computing cluster at AEI Hannover \cite{Atlas} funded by the Max Planck Society and the State of Niedersachsen, Germany. This research has made use of data, software and/or web tools obtained from the LIGO Open Science Center (\url{https://losc.ligo.org}), a service of LIGO Laboratory, the
LIGO Scientific Collaboration and the Virgo Collaboration.  LIGO is funded by the U.S. National Science Foundation. Virgo is funded by the French Centre National de Recherche Scientifique (CNRS), the Italian Istituto Nazionale della Fisica Nucleare (INFN) and the Dutch Nikhef, with contributions by Polish and Hungarian institutes.

\end{acknowledgments}

\bibliography{biblio.bib}

\end{document}